\documentclass{kluwer}    

\newdisplay{guess}{Conjecture}

\begin{document}                                                                                   
\begin{article}
\begin{opening}         
\title{\bf ROBOTIC TELESCOPES AND NETWORKS:\\
NEW TOOLS FOR EDUCATION AND SCIENCE} 
\author{Fran\c{c}ois R. QUERCI and Monique QUERCI\surname{}}  
\runningauthor{F.R. QUERCI and M. QUERCI}
\runningtitle{ROBOTIC TELESCOPES AND NETWORKS}
\institute{Observatoire Midi-Pyr\'en\'ees, 14 Avenue Edouard Belin, 31400 
Toulouse, France\\
e-mail: querci@obs-mip.fr}
\date{}

\begin{abstract}
Nowadays many telescopes around the world are automated and some networks of 
robotic telescopes are active or planned. Such equipment could be used for the
training of students and for science in the Universities of DCs and of new
astronomical countries, by sending them observational data {\it via} Internet
or through remotely controlled telescope. It seems that it is the
time to open up for discussion with UN and ESA organizations and also with
IAU, how to implement links between robotic telescopes and such
Universities applying for collaborations. Many scientific fields could
thus be accessible to them, for example on stellar variability,
near-earth object follow-up, $\gamma$-ray burst counterpart tracking, 
and so on.
\end{abstract}


\end{opening}           

\section{Introduction}  
 
The telescopes about which we talk, are mainly small-sized telescopes from
50-cm to 1-m diameters and middle-sized telescopes of the 1-m to 3-m
class. A lot of science can be done with them. Astrophysics of the next
century do not need only large telescopes of the 10-m to 40-m class, but
also small ones which can open up new fields of research also.\\
Today, our knowledge is still limited on subjects such as: 
- stellar variability (all stars are variable) studied through
asterosismology which determines the various pulsation modes of the
stellar interior, and through the modeling of stellar photospheres and
external envelopes which has to take into account the eventual presence of
spots, disks, jets, etc.,
- visible counterparts of $\gamma$-ray bursts (hereafter GRBs),
- near-earth objects (hereafter NEOs) (see the japanese proposal in these
proceedings),
- supernova (it exists supernova search programs),
- planets around stars (the first exoplanets were discovered
with the 1.93-cm telescope in {\it Haute-Provence Observatory}, France).
  
Many of these scientific objectives imply to observe the same stellar objects 
during a long time, before they can be reached.\\
Many tools were, are, and will be developed for such researches. As an
example about the stellar variability, in the past the multi-site and 
multi-wavelength observing campaigns with
current telescopes were developed for studying short-period variables
such as the white dwarfs and the $\delta$ Scuti stars, since such campaigns
could be organized only during a few weeks by year ({\it e.g.} as discussed
in Querci and Querci, 1998a).
Today, dedicated automated telescopes (AT) or robotic telescopes (RT),
or networks of them, as well as networks of interferometers are a
powerful tool for the follow-up observations of all the types of variables. 
Tomorrow, orbiting telescopes such as WSO (World Space Obs.) will challenge
the networks.\\

Coming back to the observing campaigns with current telescopes, and to AT, 
RT and Networks of them, the latter seem the best adapted equipment to 
the DCs and the new astrophysical countries at a very low cost relatively 
to the benefit they could gain, for investigating new fields of research, for 
example on object variability, and/or for preparing future observations with 
the large telescopes.\\

\section{Overview of the equipment}

Firstly, let us recall some definitions:\\
- {\bf Automated telescopes (AT)} follow a prescribed set of procedures and 
perform the indicated tasks.\\
- {\bf Robotic (or automatic) telescopes (RT)} operate without human help at
all. They offer a remote operation capability, however a {\bf fully
robotic telescope} is not supervised either locally or remotely during
its routine operation. Improvements are required to work in hostile 
environments.\\
- {\bf Networks of RT} are under the control of a main station. The data are
{\it immediately} sent to all the members of the network.\\
The main station has in charge the control of the quality of the data
and also the control of the data archives accessible via Internet.\\
The scientific programmes are defined by an international scientific board.\\

The equipment is or will be used: - in a local manual mode for the 
student training, - in a remote control mode via Internet for the
student training and for research, - and, of course, in a complete automated 
mode for research.

It includes:\\
- a weather station with sensors for clouds, rain, humidity, snow,
sand wind, high speed wind, interfaced to a site-control computer which 
drives the dome and initializes the telescope processing.\\
- an automated dome (a dome protects the equipment from humidity during
the observations),\\ 
- or a roll-off roof which, once opened, allows fast movements
of the telescope necessary in the GRB counterpart tracking and the 
monitoring of a great number of objects per night (in some sites, when
the telescope is inactive, an impenetrable dome or roof is needed because 
of the sand winds),\\
- the telescope,\\
- and its own equipment, {\it i.e.} a CCD photometer and a spectrograph, 
and a polarimeter. More specifically: - a large-field camera to search
for supernovae, optical counterparts of GRBs and NEOs, - and/or a 
middle-sized field camera for variable star studies, - a spectrograph 
with optical fibers (at low and/or high resolutions).\\
- a computer to drive the dome, the telescope and its equipment to
perform the observations by taking into account the local and sky 
constraints.\\
- a disk storage for recording the scientific observations and their
technical parameters.\\
- an Internet server to receive the observing instruction requests and to
transmit the resulting data and the technical parameters.\\
- facilities for reducing the data on line for the network members.\\

An example of operational automatic telescopes equipped with automated
photometers for high-precision photometry is described by Henry
(1999).\\

\section{\bf AT, RT and Networks around the world for research and/or
education}

At the international workshop on APTs held in 1986 as the Seventh Annual 
Fairborn IAPPP Symposium ({\it Automatic Photoelectric Telescopes}, eds. 
D.S. Hall, R.M. Genet, and B.L.  Thurston, The Fairborn Press), 
the powerful capability of making differential photoelectric photometry
automatically was highlighted. Then, year after year, meetings on the automated
astronomy showed how a lot of research programmes are best achieved
through this technique. The usefulness of APTs for education is
also unquestionable.\\
{\bf The world-wide view} of the existing and planned robotic, 
automated, and remote-controlled telescopes appears today useful: - to help 
to link such world-wide telescopes in multi-longitude networks for 
future cooperations in time series studies or in the gathering 
of time critical data, - to inform the DCs, - and to incite them to 
participate.\\ 
Also such networks are a valuable tool for education in allowing students 
to contribute to current research in astronomy ({\it e.g.} 
the MicroObservatory, see \S 3.2).\\

We endeavour to update and to complete some source lists that can be
found at the following web site addresses:\\
{\it http://www.telescope.org/rti/automated.html} (Mark Cox),\\
{\it http://gamma.bu.edu/atn/auto\_tel.html} (John Mattox),\\
{\it http://alpha.uni-sw.gwdg.de/hessman/MONET/links.html} 
(F.V. Hessman).\\
As the AT and RT have an ever-increasing success, the following lists 
are certainly not exhaustive.\\

\subsection{Telescopes dedied mainly to research}

The classification is in order of countries by longitude from West to 
East, and by telescope sites. We also note the institution(s) in charge of 
the telescopes and we give some paper references and/or PI. For each 
individual research telescope, are given: the diameter in cm, its status 
(operational: O, under construction: C, in project, in test), the main 
research topics, educational use if specially mentioned. A special list 
in the next paragraph gives the telescopes dedied to education.\\
{\bf Canada}\\
$\bullet$ Kingston (Ontario) - Queens Univ. Astronomy Research Group\\
\hspace*{0.5cm}- AT; 40; C; for education and research\\
{\bf USA}\\
$\bullet$ Buckley (Washington), Torus Observatory\\
\hspace*{0.5cm}- computer-controlled; 40; O; asteroid astrometry for the
MPC, searching for new asteroids; will be used for education for the
surrounding colleges and universities; PI: Rich Williams\\
$\bullet$ Orcas (Washington), Heron Cove Obs. / Jaimeson Science and
Engineering\\
\hspace*{0.5cm}- computer-controlled; 50; O; asteroid searches and
follow-ups\\
$\bullet$ Lick Observatory, California/ UC Berkeley Astronomy Dept.\\
\hspace*{0.5cm}- KAIT (The Katzman Automatic Imaging Telescope);
Richmond et al., 1993, 1999\\
\hspace*{0.5cm}76; O; search for supernovae, comets\\
$\bullet$ Mt. Laguna / San Diego State University, California\\
\hspace*{0.5cm}- automatic; 50; in project; a GNAT (Global Network of
Astronomical Telescopes, see \S 3.4) node\\
$\bullet$ Fairborn Observatory (Arizona), a foundation that operates
automatic telescopes (presently 13 ones from 25 to 81 cm) for various
institutions among them:\\
\underline{Tennessee State University}:\\
\hspace*{0.5cm}- APTs (Automatic Photoelectric Telescopes); Pioneer:
Louis J. Boyd; Henry, 1995, 1999\\
\hspace*{0.5cm}75 \& 80; O; sun-like stars, extrasolar planets (joint
project with Harvard-Smithsonian Center)\\
\hspace*{0.5cm}2 * 25; O; semi-regular variables\\
\hspace*{0.5cm}41; O; chromospherically active stars (joint project with
Vanderbilt University)\\
\hspace*{0.5cm}3 * 80; C\\
\hspace*{0.5cm}- AIT (Automated CCD Imaging Telescope)\\
\hspace*{0.5cm}61; C; web-based interface - authorized users will submit
their observing requests\\
\hspace*{0.5cm}- AST (Automatic Spectroscopic Telescope); Eaton, 1995\\
\hspace*{0.5cm}206; C; wide range of projects in high-resolution echelle
spectroscopy: magnetic cycles, variability, winds, pulsation in cool
stars\\
\underline{Vienna University (Austria)}\\
\hspace*{0.5cm}- APTs; {\it e.g.} Strassmeier et al., 1997a\\
\hspace*{0.5cm}twin 75 (Wolfgang-Amadeus); O; cool starspot monitoring,
asteroseismology ($\delta$ Scuti, $\lambda$ Boo stars), monitoring of AGB
stars\\ 
\underline{Four College consortium (USA)}\\
\hspace*{0.5cm}- APT, P.I.: Bob Dukes; 75; O; wide range of observing
programmes: pre-main sequence variables, pulsating variables, Be stars,
post AGB variables, solar proxies, high velocity stars, RS CVn stars, 
BY Dra variables, eclipsing binaries, etc.\\
$\bullet$ Winer Mobile Observatory (WMO), Sonoita, Arizona, provides site and
maintenance services for remotely operated telescopes through cooperative 
agreements with other astronomical institutions of which:\\
- University of Iowa (Iowa Robotic Telescope Facilities: IRTF) (see 
\S 3.2)\\
- Washington University (St. Louis, Missouri); 50; C; joint project
with the Indian Institute of Astrophysics (see India, Ladakh, below)
for GRB object monitoring\\
$\bullet$ Tenagra Observatories, Sonoita, Arizona\\
\hspace*{0.5cm}- RT; 35; O; supernova patrol; minor planet discovery;
time series studies of variable stars; also dedicated to education;
contact: M. Schwartz, Cottage Drove, Oregon\\
\hspace*{0.5cm}- RT; 81 and 50; C\\ 
$\bullet$ Kitt Peak, Arizona / SARA consortium (five Universities)\\
\hspace*{0.5cm}- SARA Telescope, remote observing possible; Oswalt et
al., 1994\\
\hspace*{0.5cm}90; O; Mira variables, Vega-type stars, white dwarfs, BL
Lac, etc.; hosts a REU (Research Experiences for Undergraduates) program
(sponsor: NSF)\\
$\bullet$ Kitt Peak, Arizona / Automatic Telescope Network (ATN)-Node\\
\hspace*{0.5cm}- AOK Telescope; robotic; Mattox et al., 1999\\
\hspace*{0.5cm}130; proposed to be a prototype ATN (see \S 3.4); optical 
counterparts of GRB; blazar monitoring (during the GLAST GR mission)\\
$\bullet$ Tucson, Arizona\\
\hspace*{0.5cm}- ARO (Apogee Robotic Observatory)\\
\hspace*{0.5cm}35; C; test bed for customers; suited for group research
and education\\
$\bullet$ Tucson, Arizona / Clemson University, South Carolina / Lawrence
Livermore National Laboratory (LLNL)\\
\hspace*{0.5cm}- Super-LOTIS (Livermore Optical Transient Imaging
System)\\
\hspace*{0.5cm}60; C; dedicated to GRB optical counterpart search\\
$\bullet$ NF/Observatory, New Mexico / Western New Mexico Univ.\\
\hspace*{0.5cm}- {\it Automatic Radio-Linked Telescope (ARLT)}; Robotic
and remote; radio-linked to Internet; Neely, 1995\\
\hspace*{0.5cm}44; O; BL Lac objects, asteroids, supernova search, Mira
survey, etc., by CCD Imaging\\
$\bullet$ Los Alamos National Laboratory, New Mexico\\
\hspace*{0.5cm}- ROTSE (Robotic Optical Transient Source Experiment);
PI: C.W. Akerlof, Univ. of Michigan\\
\hspace*{0.5cm}- ROTSE-I telephoto camera (array of four electonic
cameras on an equatorial platform); automated pointing system; O\\
\hspace*{0.5cm}- ROTSE-II telescope: robotic twin 45 cm telescopes; O\\
\hspace*{0.5cm}search for optical counterparts of GRB in coordination
with BATSE; comets, quasars\\
$\bullet$ University of Iowa\\
\hspace*{0.5cm}- One-Meter ATF - Educational Observatory (EO institute);
Robotic and remote; 100; C; CCD Imaging and Spectroscopy; deep galaxy imaging
survey; dwarf-satellite galaxy population; Embedded Infrared Sources in
the Galactic Plane; participation in the WET observations; also for
education, see below)\\
$\bullet$ Hanna City Robotic Telescope (Illinois)\\
\hspace*{0.5cm}- Remote and Robotic; Gunn and Lamb, 1997; 21; O; CCD
photome\-try of stars; light curves of eclipsing binary stars\\
$\bullet$ Goethe Link Observatories (Morgan-Monroe Station) / Indiana
University, Bloomington\\
\hspace*{0.5cm}- AT named {\it RoboScope}; Honeycutt et al., 1989; 1994\\
\hspace*{0.5cm}40; O; long-term monitoring of cataclysmic variable stars
and related objects\\
\hspace*{0.5cm}- AT named {\it SpectraBot} for both CCD imaging and 
spectroscopy; Honeycutt et al., 1993, 1998\\
\hspace*{0.5cm}125; C; long-term monitoring of time-variable
sources\\
$\bullet$ University of Michigan\\
\hspace*{0.5cm}- ROTSE-II telescope (see Los Alamos Nat. Obs., New
Mexico); C\\
$\bullet$ Nassau Astronomical Station, Ohio / Case Western Reserve Univ.
(Ohio)\\
\hspace*{0.5cm}- RT for both CCD imaging and spectrography; contact:
Earle Luck, Observatory's Director\\
\hspace*{0.5cm}90; C; will be a partner in the {\it Hands-On Universe} 
program (see \S 3.2)\\ 
{\bf Colombia}\\
$\bullet$ Bogota - Universidad Sergio Arboleda\\
\hspace*{0.5cm}- AT; 40; C; for education and research\\
{\bf Canary Islands (Spain)}\\
$\bullet$ La Palma / Liverpool John Moores University\\
\hspace*{0.5cm}- RT; Steele, 1999\\
\hspace*{0.5cm} 200; C; long-term monitoring; rapid response;
ground-based supports simultaneously to satellite observations;
small-scale surveys (see the Web site of ROBONET \S 3.4); some school
children will also be able to use the telescope to study the universe.\\\
$\bullet$ La Palma (Roque de los Muchachos) / Copenhagen Univ.(Denmark), 
RGO (Cambridge, U.K.)\\
\hspace*{0.5cm}- CAMC (Carlsberg Automatic Meridien Circle); O; contact:
L.V. Morrison, RGO\\
\hspace*{0.5cm}- Danish Str\"omgren Automatic Telescope; 50; O; contact:
J. Viggo Clausen\\
$\bullet$ Tenerife (The Teide Observatory) / Bradford University (U.K.)\\\
\hspace*{0.5cm}- RT; Director: J. Baruch\\
\hspace*{0.5cm} 30; O; devoted to search for the counterparts to gamma
burster events detected by the Compton Gamma Ray Observatory satellite;
used by school pupils in the classroom from U.K. (also experimented from
Tokyo)\\
{\bf France}\\
$\bullet$ ROSACE Automated Telescope, CNES-Toulouse, France\\
\hspace*{0.5cm}- RT; 50; in test; satellite tracking; contact: E. Cazala
-Hourcade\\
$\bullet$ Observatoire de la C\^ote d'Azur (Plateau du Calern) / Centre
d'Etude Spatiale des Rayonnements (CESR), Toulouse\\
\hspace*{0.5cm}- AT (TAROT: Rapid Action Telescope for Transient
Objects); Contact: M. Boer\\
\hspace*{0.5cm} 25; under test; detection of optical transients from
cosmic GRB\\
{\bf South Africa}\\
$\bullet$ Sutherland / South African Astronomical Observatory (SAAO)\\
\hspace*{0.5cm}- RT; 75; C\\
{\bf Italia}\\
$\bullet$ Mt. Etna / Catania Astrophysical Obs.\\
\hspace*{0.5cm}- APT; M. Rodono, Director - {\it e.g.} Strassmeier et al., 
1997b\\
\hspace*{0.5cm}80; O; RS CVn, BY Dra type stars; flare stars; chemically
peculiar A-type stars; cepheids; short-period variables\\
$\bullet$ Perugia Astronomical Observatory\\
\hspace*{0.5cm}- AIT (Automatic Imaging Telescope); Tosti et al., 1996a\\
\hspace*{0.5cm} 40; O; variability of objects brightest than 17 mag. on
V; monitoring of blazars\\
\hspace*{0.5cm}- ORIT (Optical Robotic Imaging Telescope); Tosti et al.,
1998\\
\hspace*{0.5cm}80; O; monotoring of variable sources\\ 
{\bf Slovenia}\\
$\bullet$ Crni Vrh Observatory / University of Ljubljana\\
\hspace*{0.5cm}- AT; Mikuz and Dintinjana, 1994\\
\hspace*{0.5cm} 36; O; photometry of comets; follow-up astrometric 
observations of NEOs and comets, of main-belt asteroids; variable
stars; newly discovered Novae and Supernovae; part of telescope
observing time is for advanced student observing programs.\\
{\bf Czech Republic}\\
$\bullet$ Ondrejov Robotic Telescope\\
\hspace*{0.5cm}- RT; {\it e.g.} Soldan and Nemcek, 1996\\
\hspace*{0.5cm}25; O; follow-up observations of GRB 
optical-counterparts; optical SETI observations (joint project with
Columbus Optical SETI Obs.); ground-based support of satellite
observations\\
{\bf Bulgaria}\\
$\bullet$ Belogradchik Observatory\\
\hspace*{0.5cm}- RT; Antov and Konstantinova, 1995\\
\hspace*{0.5cm}60; O; CCD photometry of fast flare events, of flare-like
events on evolved stars (red giants), of flickering of cataclysmic and
symbiotic variables, and of comets\\
{\bf Russia}\\
$\bullet$ Zvenigorod Observatory, Moscow\\
\hspace*{0.5cm}- will automate their large telescope and add it to the
TIE (Telescopes in Education) system (see \S 3.2)\\
{\bf India}\\
$\bullet$ Inter-University Centre for Astron. Astrophys. in Pune\\
\hspace*{0.5cm} Almost identical telescope to the Liverpool Telescope
(see Canary Islands, La Palma)\\
\hspace*{0.5cm}- RT; 200; in project; for use of the astronomical
community from all Indian universities\\
$\bullet$ Mt. Saraswati, Ladakh, state of Jammu and Kashmir / Indian
Institute of Astrophysics (Bangalore)\\
\hspace*{0.5cm}- AT; 50; C; joint project with Washington Univ. 
(St. Louis, Missouri); to monitor GRB objects\\
{\bf Korea}\\
$\bullet$ Yonsei University, Seoul /Center for Space Astrophys. and
Dept. of Astronomy\\
\hspace*{0.5cm}- RT; 50; O; TAOS (Taiwan-America Occultation Survey)
project\\ 
{\bf New Zealand}\\
$\bullet$ Auckland Robotic Observatory; 28; C; variable stars\\
{\bf China}\\
$\bullet$ Hong Kong / Chinese University\\
\hspace*{0.5cm}- automated; 40; C; for education and research programs\\
{\bf Taiwan}\\
$\bullet$ Taipei / Dept. of Earth Sciences, National Taiwan Normal
Univ.\\
\hspace*{0.5cm}- automated; 40; C; for education and research
programs\\ 
$\bullet$ Yu-Shan (Jade Mountain) National Park / National Central
Univ. (Inst. of Astron.) / Academia Sinica (Inst. of Earth
Sciences and Inst. of Astron. Astrophys.) / LLNL (California, USA)\\
\hspace*{0.5cm}- robotic; 3 * 50; C; TAOS project\\
{\bf Japan}\\
$\bullet$ Bisei Town, Japan Space Forum (JSF) project\\
\hspace*{0.5cm}- automated telescopes: 100 and 50 to be networked; C;  
detection of space debris, orbits of NEOs\\ 
{\bf Antarctica Plateau}\\
$\bullet$ Dome Concordia/Perugia Astronomical Observatory (Italia)\\
\hspace*{0.5cm}- IRAIT (Italian Robotic Antarctic Infrared Telescope);
Tosti et al., 1996b\\
\hspace*{0.5cm} 80 plus IR camera (8-27 $\mu$m or 8-40 $\mu$m); in project; 
long-term monitoring of variable sources;
surveys of selected sky fields at 10 $\mu$m and 20 $\mu$m; dusty environment 
in the Galaxy; old and young populations in nearby Galaxies\\

\subsection{Telescopes dedied to Education}
$\bullet$ Hawaii (Honolulu)\\
\hspace*{0.5cm}- The LCC Observatory; computer operated telescopes in
the 20-30 cm range for high school student projects; CCD imaging
systems; variability, comets; contact: Fritz Osell\\
$\bullet$ UC Santa Barbara (California)\\
\hspace*{0.5cm}- ROT (Remotely Operated Telescope), heart of RAAP
(Remote Access Astronomy Project); contact: Philip Lubin\\
\hspace*{0.5cm}35.6; O; high schools and junior colleges\\
$\bullet$ Telescopes in Education (TIE) program\\
\hspace*{0.5cm}- remotely controlled telescope at Mount Wilson
Observatory (California) used by students around the world; J. Cohen, 1997\\
\hspace*{0.5cm}60; O; widely used by schools and individuals\\
\hspace*{0.5cm}35; O; available only to select users\\
\hspace*{0.5cm}A {\bf TIE network} of world-wide automated telescopes
is in project; also TIE will restore and automate research telescopes
which will then be added to its network\\
$\bullet$ Hands-ON Universe (HOU)\\
\hspace*{0.5cm}- research programs for high and middle school
students;\\
\hspace*{0.5cm}uses LBNL (Lawrence Berkeley Nat. Obs., California) 
Nearly Supernova Search Telescope; 76; to be extended to other telescopes 
to form a network of automated telescopes for education ({\it e.g.} 
Nassau Station, see above)\\
$\bullet$ Winer Mobile Observatory (WMO), Sonoita, Arizona/ Iowa Robotic
Telescope Facilities: IRTF\\
\hspace*{0.5cm}- RT (Iowa Robotic Observatory: IRO); 50; O; primary use
is for teaching and research in undergraduate laboratories at Iowa
universities\\
$\bullet$ University of Iowa (Iowa Robotic Telescope Facilities:
IRTF)\\
\hspace*{0.5cm}- ATF (Automated Telescope Facility); 18 cm refractor; O;
primary use is for teaching and research in undergraduate laboratories
at Iowa universities\\
$\bullet$ Highland Road Park Observatory / Louisiana State Univ. (LSU) (Baton
Rouge Obs.)\\
\hspace*{0.5cm}- Remote, accessed over the Internet; connected to the
LSU Astronomy Teaching Laboratory to train teachers and students; 50;
C\\
$\bullet$ University of Wisconsin, Oshkosh\\
\hspace*{0.5cm}- {\it Stardial Number 2}, see Univ. of Illinois below; C\\
$\bullet$ University of Illinois, Urbana-Champaign\\
\hspace*{0.5cm}- {\it Stardial}, an Instructional Tool: it is an
autonomous astronomical camera for providing real-time images of the
night sky for students to access and study via the WWW; O; Dietz,
Heasley and McCullough, 1999\\
$\bullet$ CASS (Center for Automated Space Science)/NASA Univ.
Research Center, Maryland\\
\hspace*{0.5cm}- students have opportunities to participate in automated
remote observing programs with ground-based telescopes and NASA
missions\\
$\bullet$ MicroObservatory / Harvard-Smithsonian Center for
Astrophysics, New Jersey\\
\hspace*{0.5cm}- a {\bf network} of five automated telescopes controlled over
the Internet; for students and teachers nation wide; PIs: O. Gingerich
and Ph. Sadler\\
$\bullet$ Bradford Robotic Telescope Observatory on Oxenhope moor\\
\hspace*{0.5cm}- Robotic and Remote; Baruch, 1993; Cox and Baruch,
1994;\\
\hspace*{0.5cm}46; O; the telescope uses a WWW gateway to provide access
for schools, amateurs and professionals. Anyone on the Internet can register
and ask the telescope to look at anything in the northern night sky.
Students are encouraged to submit class projects.\\
\hspace*{0.5cm}Also Bradford Univ. has a 30-cm RT in Canary Islands,
Tenerife (see \S 3.1)\\
$\bullet$ Oxie, Sweden, Tycho Brahe Observatory\\
\hspace*{0.5cm}- new telescope system with complete remote-control
ability; under construction; will offer real time remote observing for
students and teachers nationwide\\

\subsection{Remote observing on research classic telescopes}
Remote observing is an observing mode where the astronomer is not
physically present at or near the telescope (see for comments, Zulstra 
et al., 1997). Robotic telescopes are a related implementation.\\
Remote observing is clearly advantageous for the many world-class large 
and giant telescopes which are in high-altitude observatories. There, 
astronomers very often feel uncomfortable and their observing 
efficiency is affected. Also, this observing mode is appropriate when 
the home institution is far from the observing site, saving astronomer's
travel and time, or when observing runs are short. Moreover, an advantage is 
that every member of a scientific cooperation can share the various
duties related to the observations and students can more easily been
included in an observing run. In fact, the imperative requirement for
remote observing is a fast data transfer system between the site
and the home. We quote:\\
$\bullet$ Mauna Kea summit, Hawaii (US)\\
\hspace*{0.5cm}- UKIRT (United Kingdom Infrared Telescope)\\
\hspace*{0.5cm}3.80 m; (Economou et al., 1996)\\
\hspace*{0.5cm}- Subaru Telescope Project (Nat. Astron. Obs. of
Japan)\\
\hspace*{0.5cm}8.3-m optical-infrared telescope; in test; will be
controlled remotely from anywhere else in the world\\
\hspace*{0.5cm}- W.M. Keck Observatory\\
\hspace*{0.5cm}twin 10-meter optical/infrared telescopes; pioneering
the routine use of remote observing: operational from Waimea, in
progress from mainland US (waiting for a reliable, separate, dedicated link;
experimented from California); Conrad et al., 1997; Kibrick, 1998\\
$\bullet$ Mauna Kea summit, Hawaii (US) and Cerro Pachon (Chile)\\
\hspace*{0.5cm}- Gemini project: two 8-m telescopes, one in Hawaii, one
in Chile; will support remote observing mode\\
$\bullet$ Hat Creek, California (US) / BIMA (Berkeley-Illinois-Maryland
Association)\\
\hspace*{0.5cm}- millimeter wave interferometer: nine 6-m diameter
antennas (Hoffman et al., 1996)\\
$\bullet$ Mt. Hopkins, F.L. Whipple Obs., Tucson, Arizona (US) /
Multiple Mirror Telescope (MMT) Observatory (CfA/OIR)\\
\hspace*{0.5cm}- conversion of the MMT in a 6.5-m telescope; C;
remote MMT observing was possible\\
$\bullet$ Kitt Peak, Tucson, Arizona (US) / NRAO\\
\hspace*{0.5cm}- 12 m; remotely observing capabilities are being
extended to a wide audience\\
$\bullet$ Apache Point Observatory, Sunspot, New Mexico (US) / Astrophysical
Research Consortium (ARC) (five Universities)\\
\hspace*{0.5cm}- 3.50 m; remote observing possible from ARC member
campuses through the Internet\\
$\bullet$ Moore Observatory, Kentucky (US) / Univ. of Louisville\\
\hspace*{0.5cm}- the Solar Flare Monitor is remotely operated; a remote
observing facility is under development on a 40-cm telescope\\
$\bullet$ La Silla (Chile) / ESO (European Southern Observatory)\\
\hspace*{0.5cm}- 1.4-m CAT (Coud\'e Auxiliary Telescope) and 3.58-m NTT
(New Technology Telescope); remote observing offered from Garching
(Germany) (Zulstra et al., 1997)\\
$\bullet$ Pico Veleta, Sierra Nevada, Granada (Spain)\\
\hspace*{0.5cm}- IRAM (Institut de RadioAstronomie Millim\'etrique)\\
\hspace*{0.5cm}30 m; remote observing at Pico Veleta possible from
IRAM Grenoble (France) and from the IRAM Granada office\\
$\bullet$ South Pole / CARA (Center for Astrophysical Research in
Antarctica)\\
\hspace*{0.5cm}- SPARO (Submillimeter Polarimeter for Antarctic Remote
Observing) operates on the Viper 2-m telescope\\

\subsection{Networks}
Since more than a decade, the idea of networks of robotic telescopes
appeared for example through the GNAT (Global Network of Automatic
Telescopes) and the NORT (Network of Oriental Robotic Telescopes) 
initially called ORT Network. Unfortunately, for various reasons 
from ventures into manufacturing robotic telescopes for GNAT  to a 
disinterestedness of the french authorities for automation in the 
case of NORT, the building of these earliest network of robotic 
telescopes for research and education has not yet come into effect. 
Today, it is an idea world-wide accepted, and the proposal of networks, 
either local or global, is growing. Comments on {\it Networks of Automated 
Telescopes Today} can be found in Williams (1999). Here are the networks 
of robotic telescopes of which we have had knowledge.\\ 
$\bullet$ {\bf ATN} (The Automatic Telescope Network) (Mattox et al.,
1999); {\it http://gamma.bu.edu/atn}\\
\hspace*{0.5cm}- CCD imaging; 8 planned telescopes of 1.3-2.0-m
diameter; GRB studies; variable stars; asteroseismology, etc.; 
partner in {\it Hands-On Universe Project} (see \S 3.2); will take over from 
the network, the Whole Earth Blazar (WEB) Telescope\\
$\bullet$ {\bf EON} (European Observational Network) for Observations of
GRB and optically violent AGNs; contact: R. Hudec, Astronomical Inst.
Ondrejov, Czech Republic; {\it http://altamira.asu.cas.cz/eon/}\\
$\bullet$ {\bf GCN} (The Gamma-ray burst Coordinates Network)
incorporated
to BACODINE (The BATSE COordinates DIstribution NEtwork); contact: S.
Barthelmy, Goddard Space Flight Center;\\
{\it http://gcn.gsfc.nasa.gov/gcn/}\\
$\bullet$ {\bf GNAT} (Global Network of Astronomical Telescopes) (Crawford,
1997); {\it http://www.gnat.org/$\sim$ida/gnat/})\\
\hspace*{0.5cm}- Imaging and photometry; a 50-cm automatic 
telescope prototype is being tested; incorporates ATIS communication 
system (see \S 3.5); gotten data are already used by graduate students
at Colorado State Univ. for their research projects\\ 
$\bullet$ {\bf GONG} (the Global Oscillation Network Group); NOAO,
Tucson, Arizona; {\it http://www.gong.noao.edu/}\\
\hspace*{0.5cm}- six-station network around the Earth (Big Bear,
Learmonth, Udaipur, Teide, Cerro Tololo, Mauna Loa) for a detailed 
study of solar internal structure through heliosismology\\
$\bullet$ {\bf LBI and VLBI networks}\\
\hspace*{0.5cm}- Centimetre wavelength Networks: EVN (European VLBI
Network), VLBA (US), Merlin, JNET (Japanese VLBI network), Australian
VLBI array, APT (Asia Pacific Telescope network)\\
\hspace*{0.5cm}- Millimetre Wavelength Network:  mm-VLBI array
(M.I.T)\\
$\bullet$ {\bf NORT} (Network of Oriental Robotic Telescopes); 
Querci and Querci, 1998a,b; 
{\it http://www.saao.ac.za/$\sim$wgssa/as2/nort.html}\\
\hspace*{0.5cm}- project selected by COPUOS, United Nations, Dec. 1996; 
6-8 stations along tropic of Cancer from Morocco to China with 1.50-m RT; 
complementary to other networks; CCD photometry, spectroscopy; variable 
stars, asteroseismology, etc.\\
$\bullet$ {\bf Robonet} A Global Network of six 2m Robotic Telescopes;
contact: Ian Halliday, Wiltshire, U.K.\\
\hspace*{0.5cm}- in project; 3 northern and 3 southern telescopes at widely
spaced longitudes; prototype: see \S3.1 Canary Islands - Liverpool John
Moores Univ.; multi-wavelength time variability, targets of opportunity, 
etc.; 
{\it http://star-www.st-and.ac.uk/$\sim$kdh1/jifpage.html}\\
$\bullet$ {\bf Spaceguard Network} (Global Network for Research on
Near-Earth Objects); joint project: Spaceguard Foundation / European
Space Agency; {\it http://www.crl.go.jp/ka/control/asteroid/SGF} \\
\hspace*{0.5cm}- automated telescopes; under study\\

\subsection{Working Groups - Softwares}
Consortiums to promote access to robotic telescopes and to develop
related softwares have been set up. Let us mention:\\ 
$\bullet$ {\bf APA} (Associate Principal Astronomer) project\\
\hspace*{0.5cm}- it performs a number of functions in support of a
human PA in planning, scheduling, and control for automatic telescopes;\\
{\it http://ic-www.arc.nasa.gov/ic/projects/xfr/}\\
$\bullet$ {\bf AstroNet}; S. Godlin and S. Chakrabarti, 1994\\
\hspace*{0.5cm}- " A toolset for simultaneous, multisite, remote
observations of astronomical objects"; project that will provide an
international network of observatories and space-based instruments\\
$\bullet$ {\bf ATIS} (Automatic Telescope Instruction Set); Henry,
1996\\
\hspace*{0.5cm}- is a language to program automatic photoelectric
telescopes; it is being improved to include spectroscopy, telescope
networking, and telescope scheduling commands\\
$\bullet$ {\bf ISTeC} (International Small Telescope Cooperative)\\
\hspace*{0.5cm}- planet-wide organization to promote access to small
($<$3 m) telescopes for research and education, and to coordinate
multi-site campaigns; {\it http://astro.fit.edu/istec}\\
$\bullet$ {\bf IWGAT} (The International Working Group on Automatic
Telescopes)\\
\hspace*{0.5cm}- is developing standards for the operation of a network
of diverse robotic telescopes to promote the development of robotic 
telescope software for general use (see ATN in \S 3.4)\\

\section{Where can we implement more such equipment?} 
An observing site research has to be done worldwide to find new
sites for the proposed networks. For example as described in Querci and
Querci (1998c, 1999), it appears that many such sites are in Developing
Countries on the basis of preliminary results on the mean annual
nebulosity from Morocco to China obtained in the frame of the NORT project.\\
The various steps in such a research are: - to prospect for 
{\bf high mountains} in semi-desertic climate (the site
has to be above the atmospheric reversing layer), - to pre-select sites 
through {\bf meteorological satellites} data, - to analyse the 
{\bf anti-correlated airstreams} for avoiding several sites on the same 
airstream, - to proceed to measurements by {\it in situ} {\bf seeing-monitor} 
or {\bf grating scale monitoring} on the pre-selected mountains, through 
campaigns done during various seasons on a few years.\\
It could be a way to introduce Astronomy and Space Science in DCs.\\

\section{How to contribute to the introduction of Astronomy and
Astrophysics, and Space Science in the DCs education cursus?}
Preliminary to going into scientific projects with robotic telescopes, some
education and training are necessary to achieve them. Let us give examples.\\
$\bullet$ Basic courses, conferences and bibliography in Astrophysics are found
on several Web sites ({\it e.g.} see also \S 3.2).\\
$\bullet$ Summer Schools are opened to the DC students to introduce
them to the basis in Astrophysics or to improve their knowledge, and
to allow them to practise a telescope often for the first time. Among
them:\\
- the IAU/UNESCO International School for Young Astronomers 
(ISYA) which is scheduled each two years since 1990,\\
- the WGSSA-UN (Working Group for the Spatial Sciences in Africa in
behalf of UN) Schools:- in South African Astrophysical Observatory for
anglophone students, - in Observatoire de Haute-Provence, south of France, for 
francophone students.\\
$\bullet$ A Newsletter within the framework of UN, dedied to Africa, 
{\it African Skies/Cieux Africains} (ed.: F.R. Querci; co-eds.: M. Querci
and P. Martinez), presents basic papers in astrophysics
and news on equipment and research from the various observatories and 
laboratories in Africa. It is freely distributed by UN and also hosted 
by the WGSSA Web page at: {\it http://www.saao.ac.za/$\sim$ wgssa/}.\\
$\bullet$ A Newsletter within the framework of the Italian Astronomical
Society presents general cultural and didactic astronomical informations
in arabic language (ed.: F. B\`onoli, Arcetri Observatory).\\

For example, to understand the stellar variability, {\bf non-stop 
observations} and hydrodynamical modelling have to be handled as shown by
the NASA/CNRS Monograph Series on {\it Nonthermal Phenomena in Stellar
Atmospheres} (eds. Stuart Jordan and Richard Thomas), for the various types
of variable stars in the H-R diagram. {\bf Indeed, the study of these objects 
does not need very large telescopes. Diameters between 50 cm and 2.00 m for 
photometry and spectroscopy respectively, are appropriate}. The non-stop 
observations are only possible with Networks of telescopes.\\
Such a subject of research could be an introduction to contemporary 
Astrophysics for the DCs.

\section{How to convince the national authorities to introduce Astronomy
and Astrophysics, and Space Science in the DCs education cursus?}

We, astrophysicists, cannot directly act on the objects that we 
study, as the chemist or the physicist does in his laboratory.
First and foremost we are observers. So, to try to understand the Universe 
that we observe, we need to develop a lot of technologies in collaboration 
with other scientists and engineers: mathematicians, physicists, 
electronics engineers, opticians, etc. Industrial challenges
are issued, first point able to arouse the national authorities's interest.
In return, the technologies developed in our observatories in
collaboration with companies can be used for the benefit of other sciences.  
Let us give an example which we have followed closely.\\
The first CCD camera were developed by army and by astronomers in
various countries. In Toulouse, in the early 80s the Observatory and
the Faculty of Medecine cooperated for the study of the coro\"{\i}de
cancer with CCDs. In mid-80s, in collaboration with the hospital of our
University, our observatory developed an experimental endoscope with a
CCD camera on its top to look inside the human body. Moreover, the CCD 
image data processing is adapted to an early identification of pathologies, 
becoming much more quantitative and accurate.\\

\section{Concluding questions}

$\bullet$ How would we coordinate the efforts made:\\
\hspace{0.5cm}- inside a new astronomical country?\\
\hspace{0.5cm}- inside a given region (Western Asia, Africa, etc.)?\\
$\bullet$ How would we open the present AT, RT, and Networks to other new
scientific groups, at least the data banks of such equipment?\\
$\bullet$ How to prepare standard training?\\
$\bullet$ How would we help out the Faculty Professors of the new astronomical
countries for their first courses in Astronomy, for their training, for
reducing the data, etc.?\\
$\bullet$ How would we help the colleagues of these countries when they plan
to develop their own equipment?\\
$\bullet$ How the {\it Northern} observatories should help out in the
achievement of the development of {\bf astrophysical laboratories} in
DCs?\\
$\bullet$ How would we convince the national authorities of these countries
to introduce Astronomy $\&$ Astrophysics and Space Sciences in their
Universities?\\
$\bullet$ How would we draw up an inventory of the needs with the help
of: - various laboratories around the World (having robotic
equipment) where DCs students are preparing PhD thesis, - IAU, - UN Outer 
Space Office?\\

In the majority of cases an adapted answer has to be found. It has to be
proposed for the benefit of science and humanity.\\

\end{article}
\end{document}